\documentclass{article}
\usepackage{epsfig,graphicx,times,amsmath,amsthm, amssymb, multicol}
\usepackage[T1]{fontenc}  
\usepackage{calc}
\usepackage{amsfonts}
\usepackage{amssymb}
\usepackage{multirow}
\usepackage{epstopdf}

\begin{document}

\title{Induced parameter-dependent optimization method  applied to reaction rate determination }
\author{Christopher G. Jesudason 
\thanks{Emails: jesu@um.edu.my or chris\_guna@yahoo.com  }\\
{\normalsize   Department of  Chemistry and Center for Theoretical and Computational Physics} \\ 
 {\normalsize Faculty of Science, University of Malaya }\\
  {\normalsize 50603 Kuala Lumpur, Malaysia} }
 \date{\today}
\maketitle


\begin{abstract} Parameter  fitting of data to a  proposed equation  almost always consider these 
parameters as independent variables. Here, the method proposed  optimizes an 
arbitrary number of variables by the minimization of a function of a single 
variable. Such a technique avoids problems associated with multiple minima and 
maxima  because of the large number of parameters, and could increase  the accuracy 
of  the determination  by cutting down on  machine errors. An algorithm for this  
optimization scheme is provided and applied  to the   determination of the rate 
constant and final concentration parameters for a first order and second order 
chemical reaction.
\end{abstract}

\section{Introduction} \label{sec:1}
Deterministic laws of nature are sometimes written - for the simplest examples- in the form 
\begin{equation} \label{eq:n1}
Y_{law}= Y_{law}(\mathbf{P},k,t)
\end{equation}
linking the variable $Y_{law}$ to $t$. The components of $\mathbf{P}$, $P_i (i=1,2,...N_p)$ and $k$ are parameters. Verification of a law of form (\ref{eq:n1})  relies on an experimental dataset $\{ (Y_{exp}(t_i), \, t_i), i=1,2,...N) \}$. Confirmation or verification of the law is based on (a) deriving suitable values for the parameters $(\mathbf{P},k)$ and (b) showing a good enough degree of fit between  the experimental set $Y_{exp}(t_i)$  and $Y_{law}(t_i)$. Many methods  \cite[etc.]{hou1,moore1,went1,went2} have been devised to determine the optimal $\mathbf{P},k$ parameters, but most if not all these methods consider the aforementioned parameters as autonomous and independent (e.g. \cite{moore1}) subjected to free and independent variation during the optimization process. On the other hand, if one considers  the interplay between  the experimental data and $Y_{law}$ one can derive certain parameters like the final concentration terms (e.g. $\lambda_\infty$ and $Y_\infty$ in what follows in sec.(\ref{sec:2}) ) if $k$, the rate constant is known. To preserve the viewpoint of the inter-relationship  between these parameters and the experimental data, we devise a scheme  that relates $\mathbf{P}$ to $k$ for all $P_i$ via  the set $\{Y_{exp}(t_i),t_i\}$, and optimize the fit over $k$-space only. i.e. there is induced a $P_i(k)$ dependency on $k$ via the the experimental set $\{ Y_{exp}(t_i),t_i\}$. it is unclear at present whether this optimization procedure is equivalent to previous ones,  but its structure  is not in contradiction  with situations where  there are inter-relations between the variables, and the results for the first and second order kinetics presented here are in very close agreement with those derived from the published literature.  the advantages  of the present method is that the optimization is over $1\,D$  $k$ space, leading to a unique determination of $\mathbf{P}$  with respect to $k$, whereas if all $\mathbf{P}$ are considered equally free, the optimization could lead to many different local solutions for each of the  $\{P_i\}$. In what follows here, we assume that the rate laws and rate constants  are not slowly varying functions of  the reactant or product concentrations, which has recently from simulation been shown generally not to be the case \cite{cgj1}. 
\section{Outline of Method} \label{sec:2}
As above, $N$ is the number of dataset pairs $\{Y_{exp}(t_i),t_i\}$, $N_p$ the number of components of the $\mathbf{P}$ parameter, and  $N_s$ the number of singularities where the use od a particular dataset $(Y_{exp}, t)$leads to a singularity in the determination of $\bar{P}_i(k)$ as defined below and which must be excluded from being used in the determination of $\bar{P}_i(k)$. Then $(N_p+1)\leq (N-N_s)$ for the unique determination of $\{\mathbf{P},k\}$. Define $^{N-N_s}C_{N_p}=N_c$ as the total number of combinations of the data-sets $\{Y_{exp}(t_i),t_i\}$ taken $N_p$ at a time that does not lead to singularities in $P_i$. Write $Y_{law}$ in the form 
\begin{equation} \label{eq:n2}
Y_{law}(t,k)=f(\mathbf{P}, t,k).
\end{equation}

Then map $f\longrightarrow Y_{th}(\bar{\mathbf{P}}, t,k)$ as follows
\begin{equation} \label{eq:n3}
Y_{th}(t,k)= f(\bar{\mathbf{P}}, t,k)
\end{equation}
where the term $\bar{\mathbf{P}} $  and its components is   defined below and where $k$ is a varying parameter.
For any of the $(i_1, i_2, \ldots, i_{N_p} )$ combinations where $i_j \equiv (Yexp(t_{i_j}) , t_{i_j})$ is a particular dataset pair, it is in principle possible to solve for the components of  $\bar{\mathbf{P}}$ in terms of $k$ through the following simultaneous equations: 

\begin{equation} \label{eq:n4}
\begin{array}{rll}
Y_{exp}(t_{i_1})&=& f(\mathbf{P}, t_{i_1} , k) \\[0.5cm]
Y_{exp}(t_{i_2})&=& f(\mathbf{P}, t_{i_2} , k) \\
 
  &\vdots&  \\
  Y_{exp}(t_{i_{N_p}})&=& f(\mathbf{P}, t_{i_{N_p}} , k)
 	\end{array}
\end{equation}
For each $P_i$, there will be $N_c$ different solutions, $P_i(k,1),P_i(k,2), \ldots P_i(k,N_c)$ . We can define (there are several possible mean definitions) an arithmetic mean for the components of $\bar{\mathbf{P}}$ as 
\begin{equation} \label{eq:n5}
\bar{P_i}(k)= \frac{1}{N_c}\sum_{i=1}^{N_c}P_i(k,j) .
\end{equation}
Each $P_i(k,j)$ is a function of $k$ whose derivative is known either analytically or by numerical differentiation.
To derive an optimized set, then for the least squares method, define 
\begin{equation} \label{eq:n5}
Q(k)= \sum_{i=1^\prime}^{N^\prime} (Yexp(t_i)-Y_{th}(k,t_i) )^2.
\end{equation}

Then for an optimized $k$, we have $Q^\prime(k)=0.$ Defining 
\begin{equation} \label{eq:n6}
P_k(k)=\sum _{i=1^\prime}^{N^\prime}(Y_{exp}(t_i)- Y_{th}(k,t_i)  ).Y_{th}^\prime(k,t_i)
\end{equation}
the optimized solution of $k$ corresponds to $P_k(k)=0.$ The most stable numerical solution is gotten by the bisection method where a solution is assured if  the  initial values of $k$ yield  opposite signs for $P_k(k).$
Since  all $\bar{P_i(k)}$ functions are known, their values  may all be computed for one optimized $k$ value of  $Q$ in (\ref{eq:n5}). For a perfect fit of $Y_{exp}$ with $Y_{law}$, $
Q(k^\prime)=Q^\prime(k^\prime)=0\Rightarrow  \bar{P_j}\rightarrow \,P_j \,\,(\forall  j) $
 and so in this sense we define the above algorithm as giving optimized values for all $P_i$ parameters via the $k$ determination. This method is illustrated for the determination of two parameters in chemical reaction rate studies, of $1^{\mbox{st} }$ and $2^{\mbox{nd} }$ order respectively using data from published literature , where this method yields values very close to those quoted in the literature. 
\section{Applications in Chemical Kinetics} \label{sec:2}
The first order reaction studied here is \\
(i) the methanolysis of ionized phenyl salicylate with data derived from the literature \cite[Table 7.1,p.381]{khan2} \\
and the second order reaction analyzed is \\
(ii)  the reaction between plutonium(VI) and iron(II) according to the data in \cite[Table II p.1427]{newt1} and \cite[Table 2-4, p.25]{bkreac15}.\\

\subsection{First order results}\label{subsec:2a}
\begin{figure}[htbp]
\begin{center}
\includegraphics[width=9cm]{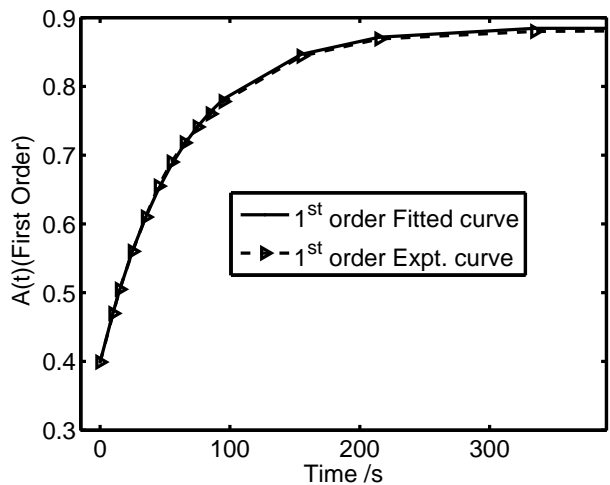} 
\end{center}
\caption{Plot of the experimental and curve with optimized parameters showing the very close fit between the two. The slight difference between the two can  probably be   attributed to experimental errors. }
\label{fig:2n} 
\end{figure}

\begin{figure}[htbp]
\begin{center}
\includegraphics[width=7cm]{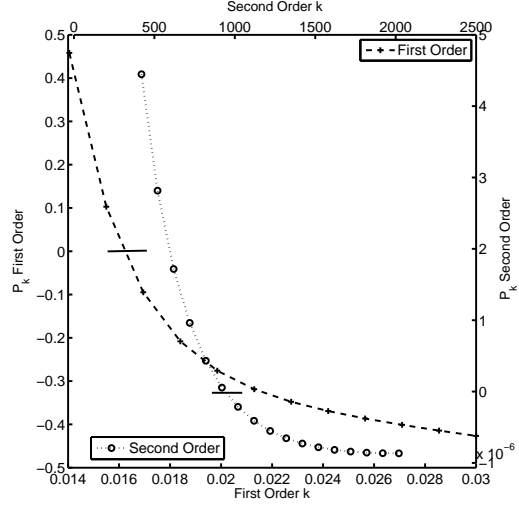} 
\end{center}
\caption{$P_k$ functions  (\ref{eq:n6})  for reactions (i) and (ii) of order one and two in reaction rate.  }
\label{fig:1n} 
\end{figure}

Reaction (i) above  corresponds to 
 \begin{equation} \label{eq:1a}
\mbox{PS}^{-}+\mbox{CH$_3$OH}\,\, \stackrel{k_a}{\longrightarrow}\,\,  \mbox{MS}^{-}  + \mbox{PhOH} 
\end{equation}
 where  the rate law is pseudo first-order  expressed as  
\[\mbox{rate}=k_a\mbox{[PS]}^{-}=k_c[\mbox{CH$_3$OH}][\mbox{PS}^{-}].\]  
with   the concentration of methanol held constant  (80\% v/v) and where the physical and thermodynamical conditions   of the reaction appears in \cite[Table 7.1,p.381]{khan2}.
The change in  time $t$ for any material property  $\lambda(t)$, which in this case  is the Absorbance $A(t)$ (i.e. $A(t)\equiv \lambda(t)$is given by 
\begin{equation} \label{eq:1}
\lambda(t)= \lambda _\infty -(\lambda_\infty - \lambda_0)\exp{(-k_at)}
\end{equation}
for a first order reaction  where $\lambda_0$ refers to the  measurable property value at time $t=0$ and $\lambda_\infty$ is the value at $t=\infty$ which is usually treated as a parameter to yield the best least squares fit  even if its optimized value is less for monotonically increasing functions (for positive $\frac{d\lambda}{dt}$at all $t$)  than an  experimentally determined $\lambda(t)$ at time $t$. In Table 7.1 of \cite{khan2} for instance, $A(t=2160 s)=0.897 >A_{opt,\infty}=0.882$  and this value of $A_\infty$ is used to derive the best  estimate of the rate constant as $16.5\pm 0.1\times 10^{-3}\mbox{sec}^{-1}$. \\
For this reaction, the $P_i$ of (\ref{eq:n2}) refers to $\lambda_\infty$  so that  $\mathbf{P}\equiv \lambda_\infty$  with $N_p=1$ and $k\equiv k_a$. To determine the parameter $\lambda_\infty$ as a function of $k_a$ according to (\ref{eq:n5}) based on the \emph{entire}  experimental $\{(\lambda_{exp},t_i)\}$ data set we invert (\ref{eq:1}) 
and write 
\begin{equation} \label{eq:1c}
\lambda_{\infty}(k)=\frac{1}{N^\prime}\sum_{i=1^\prime}^{N^\prime}\frac{(\lambda_{exp}(t_i)- \lambda_{o}\exp{-kt_i} )}{(1-\exp{-kt_i})}
\end{equation}
where the summation is for all the values of the experimental dataset that does not lead to singularities, such as when $t_i=0$, so that here $N_s=1$. We define the non-optimized, continuously deformable theoretical curve $\lambda_{th}$ where $\lambda_{th}\equiv Y_{th}(t,k)$ in (\ref{eq:n3}) as 
\begin{equation} \label{eq:1d}
\lambda_{th}(t,k)= \lambda _\infty(k) -(\lambda_\infty(k) - \lambda_0)\exp{(-k_at)}
\end{equation}
With such a projection of the $\lambda_{\infty}$ parameter $P$ onto $k$, we seek the  least square minimum of $Q_1(k)$, where $Q_1(k)\equiv Q$ of (\ref{eq:n5}) for this first-order rate constant k  in the form  
\begin{equation} \label{eq:1e}
Q_1(k)=\sum_{i=1}^N (\lambda_{exp}(t_i) -\lambda_{th}(t_i,k))^2
\end{equation}
where the summation is over all the experimental $(\lambda_{exp}(t_i), t_i)$ values. 
The resulting $P_k$ function (\ref{eq:n6})  for the first order reaction based on  the published dataset is given in Fig.(\ref{fig:1n}).The solution of the rate constant $k$ corresponds to the zero value of the function, which exists for both orders. The $\mathbf{P}$ parameters ($\lambda_\infty$ and $Y_\infty$ ) are derived by back substitution into eqs. (\ref{eq:1c}) and (\ref{eq:3}) respectively. The Newton-Raphson (NR) numerical procedure \cite[p.362]{nrc}was used to find the roots to $P_k$.For each dataset, there exists a value for $\lambda_\infty$ and so the error expressed as a standard deviation may be computed. The tolerance in accuracy for the NR procedure was $1.\times 10^{-10}$ . We define the function deviation $fd$ as the standard deviation of the  experimental results with the best fit curve $fd=\surd\frac{1}{N} \{\sum_{i=1}^N(\lambda_{exp}(t_i)-\lambda_{th}(t_i)^2\}$ Our results are as follows:\\
 $k_a=1.62\pm.09\times10^{-2}\mbox{s}^{-1}$; $\lambda_\infty=0.88665\pm.006$; and $fd=3.697\times 10^{-3}$. \\ The experimental estimates are :\\
  $k_a=1.65\pm.01\times10^{-2}\mbox{s}^{-1}$; $\lambda_\infty=0.882\pm 0.0$; and  $fd=8.563\times 10^{-3}$.\\
 The experimental method involves adjusting the $A_\infty\equiv \lambda_\infty$ to minimize the  $fd$ function and hence no estimate of the error in $A_\infty$ could be made. It is clear that our method has a lower $fd$  value and is thus a better fit, and the parameter values can be considered to coincide with the experimental estimates  within experimental error. Fig.(\ref{fig:2n})shows the close fit between the curve due to our optimization procedure and experiment. The slight variation between the two curves may well be due to experimental uncertainties.  
 
\subsection{Second order results}\label{subsec:2b}
To further test our method, we also analyze the second order reaction
 \begin{equation} \label{eq:1b}
\mbox{Pu(VI)}+2\mbox{Fe(II)}\,\, \stackrel{k_b}{\longrightarrow}\,\,  \mbox{Pu(IV)}  + 2\mbox{Fe(III)} 
\end{equation}
whose rate is given by $\mbox{rate}=k_0[\mbox{PuO}^{2+}_2][\mbox{Fe}^{2+}]$ where $k_0$ is relative to the constancy of other ions in solution such as $\mbox{H}^+$. The equations are very different in form to the first-order expressions and serves to confirm the viability of the current method.

For  Espenson, the above stoichiometry  is kinetically equivalent to the reaction scheme  \cite[eqn. (2-36)]{bkreac15}
\[\mbox{PuO}_{2}^{2+}+ \mbox{Fe}^{2+}_{aq}   \stackrel{k_b}{\longrightarrow}\, \mbox{PuO}^+_2 +\mbox{Fe}^{3+}_{aq}.\]  
which also follows from the work of  Newton et al. \cite[eqns. (8,9),p.1429]{newt1}  whose data \cite[TABLE II,p.1427]{newt1} we use and analyze to  verify  the principles presented here. Espenson had also used the same data as we have to derive the rate constant and other parameters \cite[pp.25-26]{bkreac15} which is used to  check the accuracy of our methodology. The overall absorbance in this case $Y(t)$ is given by \cite[eqn(2-35)]{bkreac15}
\begin{equation} \label{eq:2}
Y(t)= \frac{ Y_{\infty} + \left\{ Y_0 \left( 1-\alpha \right) -Y_\infty    \right\}\exp(-k\Delta_0 t)   }
{1-\alpha\exp(-k\Delta_0t)}
\end{equation}
where $\alpha=\frac{[\text{A}]_0}{[\text{B}]_0}$ is the  ratio  of initial concentrations where $[\text{B}]_0>[\text{A}]_0$ and $[\text{B}]=[\mbox{Pu(VI)}]$, $[\text{A}]=[\mbox{Fe(II)}]$ and $[\text{B}]_0=4.47\times 10^{-5}\text{M}$ and $[\text{A}]_0=3.82\times 10^{-5}\text{M}$ . A rearrangement of (\ref{eq:2}) leads to  the equivalent expression \cite[eqn(2-34)]{bkreac15}

\begin{equation} \label{eq:3}
\ln \left\{ 1+ \frac{\Delta_0\left(Y_0-Y_\infty\right)}{[\text{A}]_0\left(Y_t - Y_\infty \right)}\right\}
=\ln\frac{[\text{B}]_0}{[\text{A}]_0} +k\Delta_0t. 
\end{equation}
According to Espenson, one cannot use this equivalent form \cite[p.25]{bkreac15} "because an experimental value of $Y_\infty$ was not reported."  However, according to Espenson, if $Y_\infty$ is  determined autonomously, then $k$ the rate constant may be determined. Thus, central to all conventional methods is the autonomous and independent status of both $k$ and $Y_\infty$.  We overcome this interpretation by defining $Y_\infty$ as a function of the total experimental spectrum of $t_i$ values and $k$  by inverting (\ref{eq:2}) to define $Y_{\infty}(k)$ where 
\begin{equation} \label{eq:3b}
Y_{\infty}(k)=\frac{1}{N^\prime}\sum_{i=1^\prime}^{N^\prime}\frac{Y_{exp}(t_i)\left\{\exp(k\Delta_0t_i)-\alpha) \right\}+Y_0(\alpha-1)}{(\exp(k\Delta_0t_i)-1)}
\end{equation}
where the summation is over all experimental values that does not lead to singularities such as at $t_i=0$.
In this case, the $\mathbf{P}$ parameter is given by Y$_{\infty}(k)=P_1(k)$, $k_b=k$ is the varying $k$ parameter of (\ref{eq:n2}).  We likewise define a continuously deforming function $Y_{th}$ of $k$ as
\begin{equation} \label{eq:3c}
Y(t)_{th}= \frac{ Y_{\infty}(k) + \left\{ Y_0 \left( 1-\alpha \right) -Y_\infty(k)    \right\}\exp(-k\Delta_0 t)   }
{1-\alpha\exp(-k\Delta_0t)}
\end{equation}
 In order to extract the parameters $k$ and $Y_{\infty}$ we minimize
 the  square function   $Q_2(k)$  for this second order  rate constant with respect to $k$  given as 
\begin{equation} \label{eq:3d}
Q_2(k)=\sum_{i=1}^N (Y_{exp}(t_i) -Y_{th}(t_i,k))^2
\end{equation}
where the summation are over the experiment $t_i$ coordinates. Then the solution to the minimization problem is when the corresponding $P_k$ function (\ref{eq:n6}) is zero. The NR method was  used to  solve $P_k=0$ with the error tolerance of $1.0\times10^{-10}$.
With the same notation as in the first order case, the second order results are:\\
  $k_b=938.0\pm 18 \mbox{M s}^{-1}$; $Y_\infty=0.0245 \pm 0.003$; and $fd=9.606\times 10^{-4}$. \\
 
\begin{figure}[htbp]
\begin{center}
\includegraphics[width=9cm]{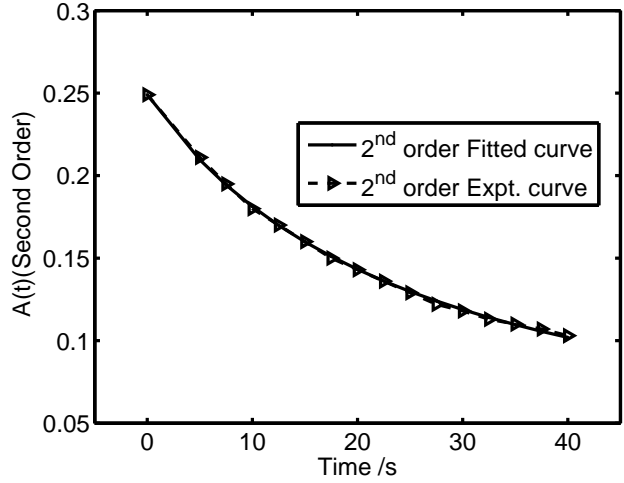} 
\end{center}
\caption{Graph of the experimental and calculated curve based on the current induced parameter-dependent optimization method. }
\label{fig:3n} 
\end{figure}

The experimental estimates are \cite[p.25]{bkreac15}:\\
  $k_b=949.0\pm 22\times10^{-2}\mbox{s}^{-1}$; $Y_\infty=0.025 \pm 0.003$.\\
  Again the two results are in close agreement. The graph of the experimental curve and the one that derives from our optimization method in given in Fig.(\ref{fig:3n}).
  
\section{Conclusions}
The results presented here show that for linked variables, it is possible to derive all the parameters associated with a curve by considering only one independent variable which serves as a function of all the other variables in the optimization process that uses experimental dataset as input variables in the estimation. Apart from possible reduced errors in  the computations, there might also be  a more accurate way of deriving parameters that are more determined by the value of one parameter (such as $k$ here) than others; the current methods that gives equal weight  to all the variables might in some cases lead to results that would be considered "unphysical".

\section{Acknowledgments}
This work was supported by University of Malaya Grant UMRG(RG077/09AFR) and Malaysian Government grant   FRGS(FP084/2010A).  \\ \\
 
 \bibliographystyle{unsrt}	
\bibliography{mpbib}

\end{document}